\documentclass[letterpaper]{article}
\pdfoutput=1
\usepackage{color}
\usepackage{float}
\usepackage{tabularx,colortbl}
\usepackage{setspace}
\usepackage{cite,graphicx,amsmath,psfrag,bm}
\usepackage{subfigure}
\usepackage[usenames,dvipsnames]{xcolor}
\usepackage{mathabx}
\usepackage{cancel}
\usepackage{epstopdf}
\usepackage[T1]{fontenc}
\usepackage[utf8]{inputenc}
\usepackage{authblk}
\usepackage[process=auto]{pstool}
\usepackage[normalem]{ulem}
\graphicspath{{Figures/}}

\title{Metasurface Particle with Independent Transmission and Reflection Full Phase Coverage}
\author{Ashutosh Patri\thanks{ashutosh.patri@polymtl.ca},  Guillaume Lavigne,  Nima Chamanara, and Christophe Caloz }
\affil{Department of Electrical Engineering, Polytechnique Montr\'{e}al\\}

\begin{document}
\maketitle
\begin{abstract}
A metasurface particle with independent transmission and reflection full phase coverage for circularly polarized waves is introduced. This particle is constituted of two parts, one controlling the power splitting and the reflection phase, and the other one controlling the transmission phase, both leveraging the Pancharatnam-Berry phase principle. Given its unique flexibility, this particle may find various applications in metasurface technology.
\end{abstract}

\section{Introduction}
Metasurfaces are 2D metamaterials capable of transforming electromagnetic waves with an unprecedented accuracy and diversity. Sophisticated synthesis methods have recently been developed for the determination of the surface susceptibility tensor functions of these metasurfaces~\cite{achouri2015synthesis, achouri2017design}. However, this operation constitutes only the first step of the complete synthesis procedure, the second step being the determination of the corresponding scattering particles. This second step is still very challenging. For example, metasurface holographic imaging typically requires sophisticated particles with full phase and amplitude coverage, and independent phase and magnitude control is generally hard to achieve.

The Pancharatnam-Berry phase concept has recently helped to solve this issue to some extent for the case of circularly polarized (CP) light~\cite{kang2012wave}. In the application of this concept, a \emph{unique particle shape}, acting as a half-wave plate element, provides \emph{full phase coverage} with constant amplitude upon simple geometrical rotation, which considerably simplifies the design of the metasurface.

Until recently, this Pancharatnam-Berry particle design technique was restricted to \emph{transmission only} or \emph{reflection only} designs. In~\cite{khan2017simultaneous}, this restriction has been lifted, but with a method leading to \emph{dependent reflection and transmission} phases. This dependence represents a limitation in most applications, including that of a hologram producing different images in reflection and transmission, which we are currently working on.

In order to eliminate this limitation, we introduce here a novel Pancharatnam-Berry metasurface particle design providing \emph{independent reflection and transmission full phase coverage} for CP waves.

\section{Particle Design Rationale}
\subsection{General Idea}

\begin{figure}[h]
\centering

\includegraphics*[width=3in]{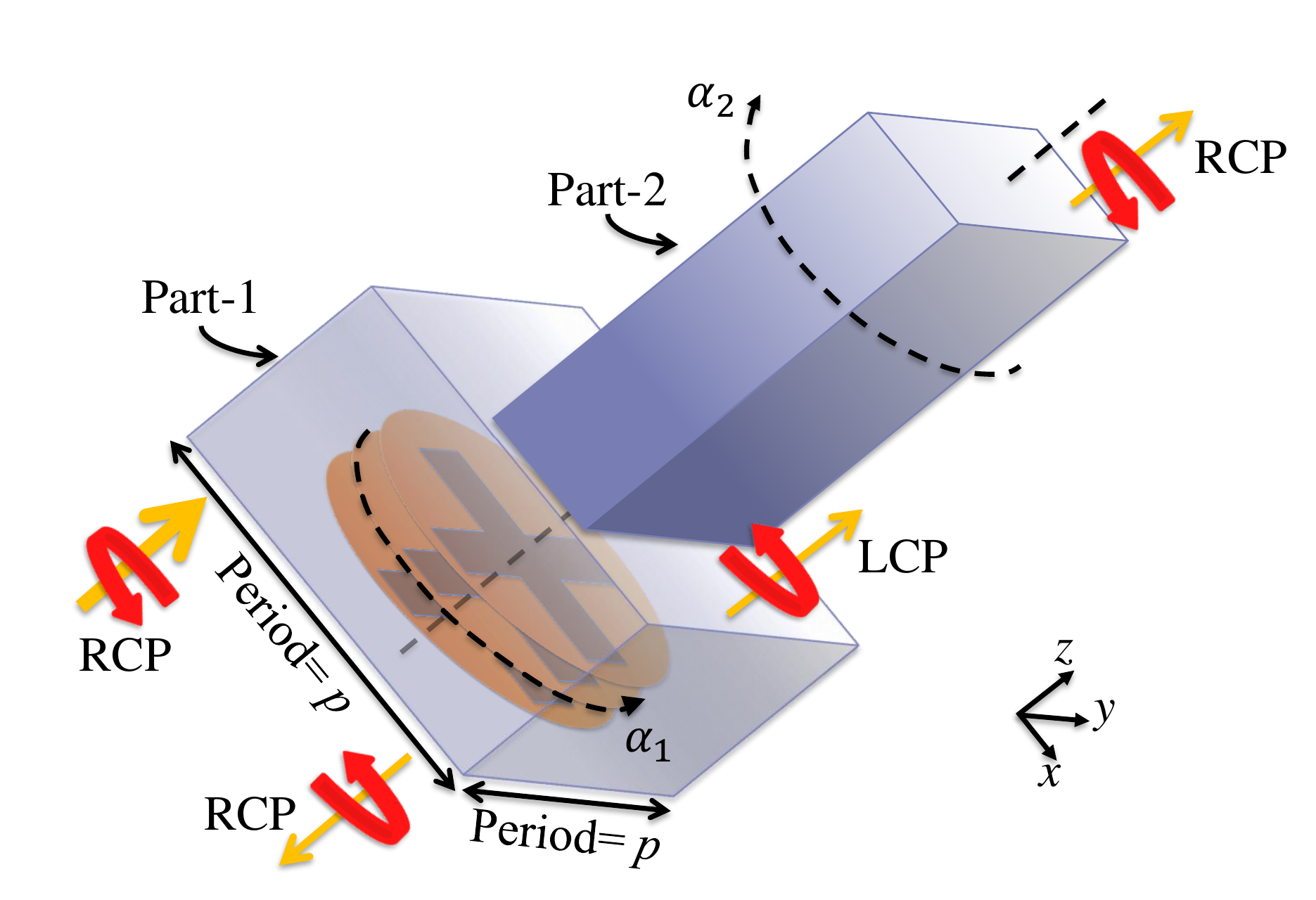}
\caption{Proposed metasurface particle for independent transmission and reflection full phase coverage using the Pancharatnam-Berry principle, which assumes CP waves. Full phase coverage is achieved by tuning the angle pair $(\alpha_1,\alpha_2)$.}
\vspace{-0.1cm}
\label{meta_particle_3D_view}
\end{figure}

The proposed metasurface particle is shown Fig.~\ref{meta_particle_3D_view}. Its design rationale is composed of two steps, corresponding to the two particle-parts as indicated in the figure, both leveraging the Pancharatnam-Berry phase principle. Part-1 is a metallo-dielectric half-wave plate half mirror controlling the transmission and reflection phase and magnitude responses dependently, while Part-2 is a reflection-less half-wave plate allowing independent control of the transmission phase. So, the key idea is to first design Part-1 for the desired reflection phase, without paying attention to the transmission phase, and to next adjust the overall transmission by tuning Part-2, which does not have any impact on reflection. 

\begin{figure}[h!]
\begin{center}
\centering
\vspace{-0.4cm}
\includegraphics*[width=3in]{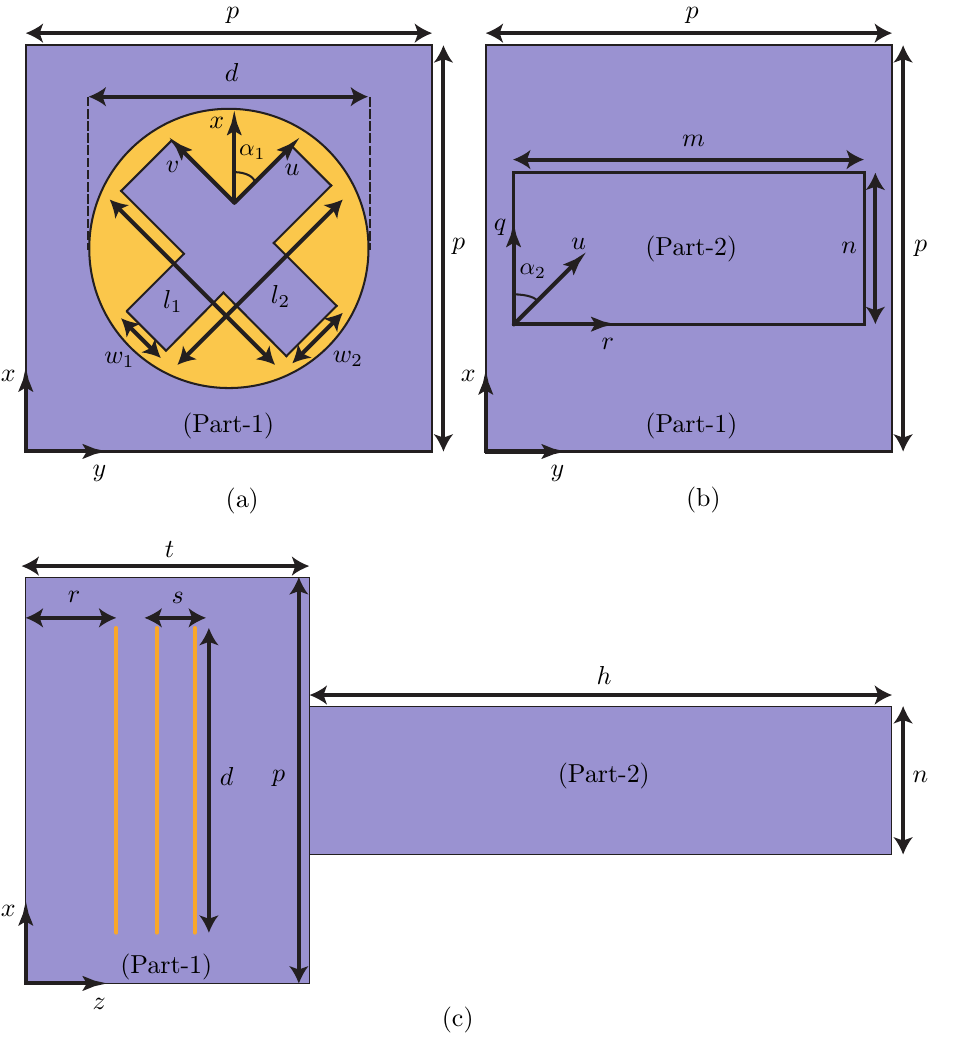}
 \caption{Design parameters and initial values for half-power splitting for the proposed metasurface cell in Fig.~\ref{meta_particle_3D_view}. (a)~Bottom $xy$-view. (b)~Top $xy$-view. (c)~Side $xz$-view. The parameters values in the forthcoming design example are, in mm, $p=10$, $d=9$, $l_{1}=8$, $l_{2}=5.25$, $w_{1}=1.3$, $w_{2}=1.6$, $m=8.4$, $n=4.2$, $h=53$, $t=18.24$, $s=1.52$ and $r=7.6$, for a particle with desired properties at 10.7 GHz. The Pancharatnam-Berry key parameters are the rotations angles $\alpha_1$ and $\alpha_2$, that have here been taken arbitrarily.
  }\label{meta_particle_dimension}
\end{center}
\vspace{-0.7cm}
\end{figure}

The design parameters and initial values for the particle in Fig.~\ref{meta_particle_3D_view} are given in Fig.~\ref{meta_particle_dimension}. The initial values are the parameter values that have been optimized (FEM CST Microwave Studio) for half power splitting. In this optimization, the parameters $\alpha_1$ and $\alpha_2$ have been taken arbitrarily, as they do not affect the magnitude; they will be later the tuning parameters for full-phase coverage.

\subsection{Part-1 Design}

Part-1 is intended to control the overall magnitude response and is hence designed to perform the half power separation between the reflected and transmitted wave, in view of our forthcoming double-image hologram application. It is therefore characterized by the transmission and reflection coefficient magnitudes
\begin{equation}
|T_{x}|^2=|T_{y}|^2=|R_{x}|^2=|R_{y}|^2=0.5,\label{EQ1}
\end{equation}
where the subscripts indicate the polarization. Moreover, \mbox{Part-1} is designed as a half-wave plate element in order to leverage the Pancharatnam-Berry principle, i.e.
\begin{equation}
\Delta\phi(R_{x},R_{y})\equiv\Delta\phi(T_{x},T_{y})=\pi,\label{EQ2}
\end{equation}
where it appears that the CP handedness is reversed ($\pi/2\rightarrow 3\pi/2$) for the transmission.

Part-1 [Figs.~\ref{meta_particle_3D_view} and \ref{meta_particle_dimension}] of the particle includes three layers~\cite{pfeiffer2013cascaded} of equally spaced circular metallic discs hosting each a cross-shaped slot. These layers are embedded in a dielectric block of $\epsilon_\text{r}=6$. According to the Pancharatnam-Berry effect, the simultaneous rotation of the three discs by a \emph{geometrical angle} $\alpha_{1}$ leads to the \emph{phase shift} of $2\alpha_{1}$ with CP handedness reversal in transmission. The choice of the circular shape for the metallizations supporting the slots allows minimal coupling variation among adjacent particles upon rotation, which stabilizes the transmission and reflection amplitude over the full phase coverage.

\subsection{Part-2 Design}

As mentioned above, Part-2 is reflection-less and its only role is to adjust the transmission phase. The coupling between Part-1 and Part-2 should therefore be minimal so that the addition of the latter to the former does not alter its response. This condition is ideally fulfilled using a matched rectangular dielectric waveguide as Part-2. Again, to leverage the Pancharatnam-Berry effect, this waveguide is designed as a half-wave plate element, but of course with full transmission. The total transmitted phase amounts then to $\Phi_{T_\text{CP}}=2\alpha_{1}+2\alpha_{2}$. It should be noted that the orientation angle $\alpha_{2}$ is opposite to that of $\alpha_{1}$, because the incident wave for Part-2 has the opposite handedness of the incident wave for Part-1.

\section{Simulation Results}

\begin{figure}[h]
\begin{center}
\centering
\vspace{-0.4cm}
\includegraphics*[width=3in]{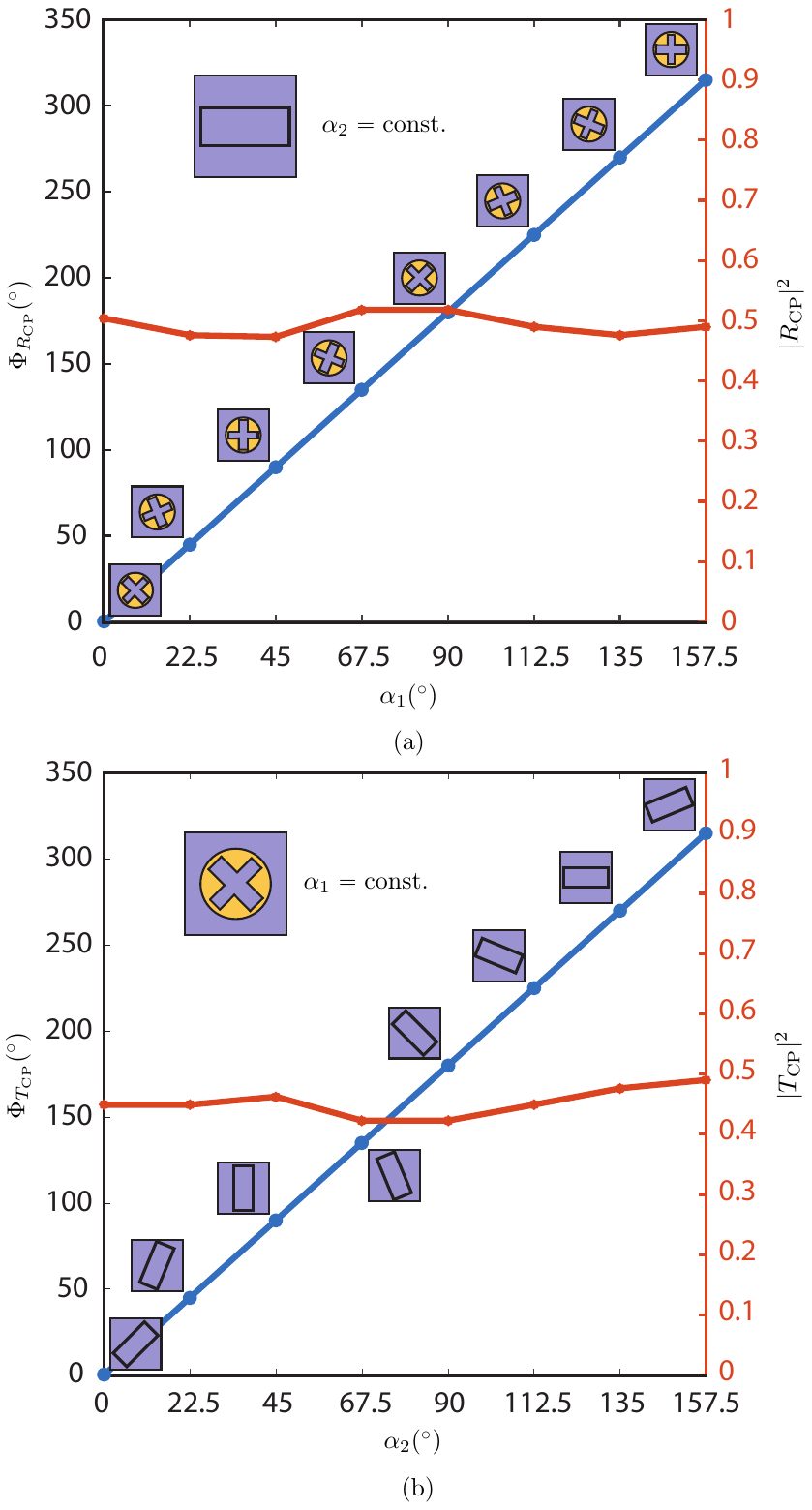}
\caption{Full-wave simulated response of metasurface particle in Fig.~\ref{meta_particle_dimension}. (a) Reflected phase and reflectance versus $\alpha_{1}$. (b) Transmitted phase and transmittance versus $\alpha_{2}$.}\label{meta_particle_simulation_results}\vspace{-3 mm}
\end{center}
\end{figure}

Figure~\ref{meta_particle_simulation_results} presents the full-wave simulation (FEM CST Microwave Studio) results of the proposed particle. It demonstrates that the proposed metasurface particle indeed provides independent reflection and transmission full phase coverage.

\section{Conclusion}

A metasurface particle for independent transmission and reflection full phase coverage has been introduced. This particle may find applications in metasurfaces with different transmission and reflection specifications, such as hologram with different reflection and transmission images.
%\nocite{*}

\end{document}